
 \magnification=1200
\headline{\ifnum\pageno=0 \nopagenumbers
\else \hss\number \pageno \fi}
\vsize=22truecm
\hsize=15truecm

\overfullrule=0pt
\hoffset=-2truecm
\font\boldgreek=cmmib10 \textfont9=\boldgreek
\mathchardef\mymu="0916 
\footline={\hfil} \hoffset=2truecm \voffset=-14pt \hsize=14.0truecm
\vsize=25.0truecm \baselineskip=20pt 
\voffset 2cm
\vbox to 1,5truecm{}
 \parskip=0.2truecm \centerline{\bf Planck Scale Effects}
\medskip
 \centerline{\bf and Axions in Supersymmetry}\bigskip
\vskip 1 truecm \centerline{by}\bigskip
\centerline{ E.A.Dudas\footnote{*}{e.mail dudas@qcd.ups.circe.fr}}
\medskip
 \centerline{\it Laboratoire de Physique Th\'eorique et Hautes Energies, *
\footnote{*}{*Laboratoire associ\'e au Centre National de la Recherche
Scientifique}}  \centerline{\it Universit\'e de Paris XI, b\^atiment 211, 91405
Orsay
Cedex, France}
\vskip 0.5cm
\centerline{ABSTRACT}
\vskip 0.5cm
  The effects of possible explicit violation of the Peccei-Quinn symmetry
responsible for the solution of the strong CP problem are studied in
supersymmetric models. It is shown that automatic models with an abelian
$U(1)$ gauge symmetry are easy to construct both in the context of
fundamental and composite models of axions. It is argued that it is
preferable to use abelian rather that nonabelian gauge groups in order
to obtains automatic symmetries. A composite model with no exact
R symmetry  is studied and it is found that, unlike common belief,
supersymmetry is broken.

\vskip 2cm
\noindent LPTHE Orsay 93/35 \par
\noindent September 1993
\vfill\supereject
\vskip 1cm
One of the open problems of the standard model arises from nonperturbative
effects in the QCD sector. Essentially, QCD instantons induce a term
$$\theta {{\rm g}^2 \over 32 \pi^2} {\rm F}^{a \mu \nu} \ {\rm F^{*}}_{a \mu
\nu}$$

\noindent in the effective lagrangian which violates both P and CP symmetries
[1]. As a consequence a neutron electric-dipole moment of order ${\rm d_n}
\simeq
5.10^{-16} \theta \ {\rm ecm}$ will be induced which compared to the
experimental measurements constrains $\bar{\theta}$ to be less than $10^{-9}$.
Here $\bar{\theta} = \theta + {\rm arg} \ {\rm det} \ {\rm M_q}$, whose
${\rm M_q}$ is the fermion mass matrix coming from the Higgs-fermion Yukawa
interaction. The so-called strong CP problem is that there is no reason that
$\bar{\theta}$ fine-tune to zero to the required accuracy. \par
The candidate solutions are generally of two types. The first use Peccei-Quinn
(PQ) type models [2] with an extra global ${\rm U}(1)_{\rm PQ}$ symmetry that
is spontaneously broken at a scale $\Lambda_{\rm PQ}$ giving rise to a
Nambu-Goldstone boson $a$ known as the axion [3]. This symmetry is explicitly
broken by instanton effects which nonperturbatively generate an axion
potential minimized by $\bar{\theta} = 0$. An axion mass of order ${\rm m_a}
\sim
{\Lambda_{\rm QCD}^2 \over \Lambda_{\rm PQ}}$ is generated which, combined
with astrophysical and cosmological considerations gives us $10^9 \leq
\Lambda_{\rm PQ} \leq 10^{12} {\rm GeV}$ [4]. Models along this line have
been constructed with the invisible axion as a fundamental scalar particle
[5] or a composite fermion-antifermion bound state [6]. \par
A second candidate solution for the strong CP problem is given by the natural
models [7]. There, CP is either explicitly (hard if complex Yukawa
couplings are introduced and soft if complex scalar masses are allowed)
or spontaneously broken. \par
Recently, the Peccei-Quinn mechanism was questionned on the basis of
possible higher-dimensional operators which could explicitly violate the
Peccei-Quinn symmetry at the Planck scale $M_p$[9-10-12]. Gravitational
effects like black holes or wormholes could violate in principle any
global symmetry not protected by some gauge group. Even if these
higher-dimensional operators are suppressed by inverse powers of the
Planck scale, the fact that the Peccei-Quinn scale is not very far away
generates contributions for the axion effective potential. Adding
these terms to the usual one coming from the QCD color anomaly
results in a vacuum with $\bar{\theta} \not= \theta$. \par
Evading models where constructed in ref. [9] and [11] using the notion of
automatic symmetries [12]. In this case the gauge structure protects
the appearance of low-dimension operators breaking the Peccei-Quinn
symmetry which appears as an accidental consequence of the gauge
symmetry. \par
The purpose of this note is to investigate the Planck scale effects in
connection with the PQ solution in supersymmetric models. Solutions
along the lines of ref. [10] and [11] will be analysed in order to find
an example of a reasonable gauge group and chiral matter content
leading to the required automatic symmetry. A simple solution is found
to be a protecting U(1) gauge group with abelian charges which
automatically forbid low-dimensional PQ breaking operator in the
superpotential. Finally some consequences of supersymmetry breaking in
a composite model are given. \par
A simple remark would be that we cannot identify the PQ symmetry with
an $R$ symmetry because of terms of type  ${1 \over {M_p}^{3n-3}} \int
d^2 \theta tr (W^{\alpha} W_{\alpha} )^n$ which cannot be avoided.
So axial transformations commuting with supersymmetry will be the
only possible candidates as automatic symmetries.
A supersymmetric model using an axial PQ symmetry and no exact
R symmetry will be constructed,with fermion condensation breaking
supersymmetry. This contradicts a general result [15] which states
that a necessary condition for supersymmetry breaking is the existence
of a nonanomalous continuous $R$ symmetry which is spontaneously
broken.
\par
There are two points which make the analysis different with respect to
the non supersymmetric theories: \par
\hskip 1 truecm - The first and the most important is the structure of
the scalar potential in a global supersymmetric theory [13] which can
be written as
$${\rm V(z_i)} = \sum_{\rm i} \left | {\partial {\rm W} \over
\partial {\rm z_i}} \right|^2 + {1 \over 2} \sum_a ({\rm D_a})^2 +
{\rm V_{soft}} \ \ \ . \eqno(1)$$

\noindent In (1) ${\rm W}(\phi_{\rm i})$ is the superpotential
constructed out of the chiral superfields $\phi_{\rm i}$ which
contain the scalar complex fields ${\rm z_i}$. ${\rm D_a}$ are the
auxiliary components of the real gauge superfields and ${\rm
V_{soft}}$ contains terms breaking softly supersymmetry. \par
Breaking supersymmetry at the Planck scale is not a welcomed
possibility because a successful phenomenology and  the
hierarchy problem suggest the superpartners masses to be around 1
TeV [14]. Then more favoured possibilities seem to be the
breaking at an intermediate scale or at low-energies. Hence we
will allow only supersymmetric PQ violating terms in the
potential (1) which come from the superpotential W (in global
supersymmetry the Kahler function do not contribute to the scalar
potential). Moreover, if W consists of two terms with different PQ charges
$$ {\rm W} = {\rm W}_1 + {\rm W}_2$$

\noindent then it is clear that its contribution
$$\sum_{\rm i} \left | {\partial {\rm W}_1 \over \partial {\rm z}_i} +
{\partial
{\rm W}_2 \over \partial {\rm z}_i} \right |^2 \eqno(2)$$

\noindent to V may violate the PQ symmetry only through the
\underbar{interference terms} in (2). By definition the automatic
models have a tree level renormalisable superpotential necessary
in order to accidentally generate the PQ symmetry. Then in order
to avoid  symmetry breaking operators with dimension less than,
 say 12, we must forbid in the superpotential terms with
dimensions less than 11. Take for example the supersymmetric GUT
model given in the ref. [9] based on the gauge group ${\rm E}_6 \times {\rm
U(1)_X}$. The superfield content is several 27's with X charges
$\pm 1$ and a $\overline{351}$ with X charge 0. The only
renormalisable and gauge invariant term which can be written in
the superpotential is of the form $27_1.27_{-1}.\overline{351}_0$.
This automatically gives rise to a PQ symmetry with PQ charges
+ 1 for the 27's and - 2 for the $\overline{351}$. The lowest
dimension PQ symmetry breaking operators in the superpotential
consistent with gauge invariance are the terms
 $27_1^3.27_{-1}^3$,  $\overline{351}^6$ and
$27_1.27_{-1}.(\overline{351}_0)^4$. The interference with the tree
level renormalisable term gives dimension 7 operators breaking
PQ symmetry which spoil the PQ solution. A simple modification
of this model which do not have this problem is adding a
supplementary chiral superfield multiplet ${{\rm r}_{\rm X}}_4$
and impose the following system of equations on the ${\rm
U(1)_X}$ charges :
$${\rm Tr \ X} = 27({\rm X}_1 + {\rm X}_2) + 351 {\rm X}_3 +
{\rm r} \ {\rm X}_4 = 0$$
$${\rm Tr} \ {\rm X}^3 = 27({\rm X}_1^3 + {\rm X}_2^3) + 351
{\rm X}_3^3 + {\rm r} {\rm X}_4^3 = 0$$
$${\rm X}_1 + {\rm X}_2 + {\rm X}_3 = 0 \ \ \ .\eqno(3)$$

The first two are the conditions of anomaly cancellation for
the ${\rm U(1)_X}$ gauge group and the third one allows the
construction of the same previous superpotential $27_1 27_{-1}
\overline{351}_0$. Taking ${\rm X}_4 = 0$ the unique solution is
${\rm X}_1 + {\rm X}_2 = 0$. That is why we must consider at
least one supplementary chiral superfield multiplet. \par
As long as ${\rm dim \ r} > 351$ we have real solutions for the
system with ${\rm X}_1 + {\rm X}_2 \not= 0$ and no dangerous PQ
breaking operators can be constructed. In fact imposing a
supplementary term in the superpotential requires the
supplementary equation
$${\rm a \ X_1} + {\rm b \ X_2} + {\rm c \ X_3} + {\rm d \ X_4} = 0
\eqno(4)$$

\noindent with a, b, c, and d positive coefficients or zero and
the system (3) + (4) has generally only the trivial solution.
It can be explicitly checked anyway for $r = 650$ and $r = 1728$.
We do not allow a v.e.v. for the representation $r$ and we
assume soft breaking terms such that all the corresponding particles
will be superheavy of order the unification scale ${\Lambda}_{GUT}$
and will not contribute to the running of $\alpha_3$ between
${\Lambda}_{GUT}$ and ${\Lambda}_{QCD}$.

\par
\hskip 1 truecm - The second point is to check that the infrared
confinement of QCD is not destroyed, a constraint especially for
the composite axion models. We will take as example the model
proposed in ref. [11] and try to supersymmetrize it. The gauge
group is SU(N) $\times$ SU(m) $\times$ G, with G the standard
model gauge group. SU(N) with N $>$ 3 has a coupling constant
which becomes strong at the intermediate scale $\Lambda_{\rm
PQ}$ and SU(m) is introduced in order to protect the
low-dimension symmetry breaking operators. The supplementary fermions are
left-handed transforming under SU(N) $\times$ SU(m) $\times {\rm SU}(3)_{\rm
c}$
as
$$({\rm N, m}, 3) + 3({\rm N, \bar m}, 1) + {\rm m} ({\rm \bar N}, 1, \bar 3)
+ 3{\rm m}(\bar{\rm N}, 1, 1) \ \ \ .  \eqno(5)$$

Under the PQ symmetry the coloured fermions have charge + 1 and the
color neutral
fermions - 1. The lowest dimensional operator consistent with the gauge and
Lorentz invariance which violates the PQ symmetry is the operator with 2m color
neutral fermions, half of them N's and half $\bar{\rm N}$'s. For m $\geq$ 4 the
symmetry breaking effects are sufficiently suppressed and the PQ mechanism
still works. \par
In the supersymmetrized version of the model the representations given in eq.
(5) describe chiral superfields. Denote by $\Lambda_{\rm GUT}$ the energy where
 supplementary superheavy fields come into play. Computing the running of
$\alpha_3$ between $\Lambda_{\rm GUT}$ and an arbitrary scale $\mu <
\Lambda_{\rm PQ}$ and taking for simplicity a step-type decoupling of the heavy
fields we obtain
$${1 \over \alpha_3(\mu)} = {1 \over \alpha_3(\Lambda_{\rm GUT})} - {1 \over 2
\pi} (3 - {\rm mN}) \ell n {\Lambda_{\rm GUT} \over \Lambda_{\rm PQ}} - {3
\over
2 \pi} \ell n {\Lambda_{\rm PQ} \over \mu} \ \ \ .  \eqno(6)$$

where in the right-hand side we included the contribution of the usual
quark superfields.

\noindent Take interesting values for $\Lambda_{\rm PQ} \sim 10^7-10^{12}$ GeV
and $\Lambda_{\rm GUT} \sim 10^{15}-10^{18}$ GeV, for N $>$ m $\geq$ 4. Then
using a perturbative value for $\alpha_3(\Lambda_{\rm GUT})$ we are not allowed
to get a strong coupled QCD at low energies, because the coloured exotic
matter fields tends to decrease $\alpha_3$ above $\Lambda_{\rm PQ}$ such
that ${\alpha_3}({\Lambda}_{QCD})< {\alpha_3}({\Lambda}_{GUT})$. \par
The problems with the unmodified version of the first model ${\rm E}_6 \times
{\rm U(1)_X}$ was that the
PQ was not sufficiently protected and with the second composite model that
protecting it with a non abelian gauge group SU(m) we lost the infrared
confinement of QCD.  \par
A simple way to protect the PQ symmetry without affecting QCD is to use a
SU(N) $\times {\rm U(1)_X}$ gauge group and matter multiplets with abelian
charges such that an appropriate automatic symmetry naturally
arises. Probably it is not the only way to construct models with
suppressed Planck scale effects, but a very simple one. We will
consider a composite model in the spirit of ref.[11] and ask for the
simultaneous breaking of supersymmetry and PQ symmetry at the scale
$\Lambda_{\rm PQ}$. \par
The  chiral  superfields  content  of  the  model  transforms
 under  $SU(N)\times{ SU(3)_c}\times{ U(1)_X}$ as $$\phi_1 ({{\rm N},1 \
1)_{\rm X}}_1  + \phi_2 ({{\rm \bar N},1 \ 1)_{\rm X}}_2 + \phi_3 ({{\rm N}, \
3)_{\rm X}}_3 + \phi_4 ({{\rm \bar N}, \ \bar 3)_{\rm X}}_4 + {\rm
S}_1{(1,1)_{\rm X}}_5 +{\rm S}_6{(1,1)_{\rm X}}_6   \eqno(7)$$

\noindent where ${\rm X}_{\rm i}$ are the abelian charges. At $\Lambda_{\rm
QCD}$
condensates of type $<\psi_1 \psi_2>$ and $<\psi_3 \psi_4>$  are formed
breaking
${\rm U(1)}_{\rm PQ}$ and supersymmetry. \par
In the globally supersymmetric case in most cases the fermion condensation
does not breaks supersymmetry [14]. A necessary (but not sufficient)
condition is the existence of a non-anomalous continuous $R$ symmetry
which is spontaneously broken [15]. In the global case with a trivial
Kahler potential the present model has such a symmetry. More interesting is the
  case
with the superpotential $W$ and the Kahler potential defined below in
eq.(9) which do not have it and still supersymmetry is broken. \par
Imposing the ${\rm U(1)}_{\rm X}$ anomaly cancellation condition we will
obtain two equations for ${\rm X}_{\rm i}$. Another one is obtained
by imposing the
existence of a term ${\rm S}_1 \phi_1 \phi_2$ in the superpotential useful in
defining the automatic PQ symmetry . A fourth equation, needed for the
dynamical breaking of supersymmetry comes by imposing a term ${\rm S}_2^+
\phi_3 \phi_4$ in the Kahler potential K . \par
Hence we arrive at the following system of equations
$$ \matrix{
{\rm N(X_1 + X_2)} + 3 {\rm N(X_3 + X_4)} + {\rm X}_5 + {\rm X}_6 = 0 \cr
{\rm N(X_1^3 + X_2^3)} + 3 {\rm N(X_3^3 + X_4^3)} + {\rm X}_5^3 + {\rm X}_6^3 =
0 \cr
{\rm X}_1 + {\rm X}_2 + {\rm X}_5 = 0 \cr
{\rm X}_3 + {\rm X}_4 + {\rm X}_6 = 0 \cr
}  . \eqno(8)$$
which always has nontrivial real solutions.

\noindent Write the lowest dimensional terms allowed by the gauge symmetry
for N $\geq$ 4 (but N $\not=$ 5 where accidentally we can construct a
dimension 6 operator breaking PQ symmetry in W)
$${\rm W} = \lambda_1 {\rm S}_1 \phi_1 \phi_2 + {\lambda_2 \over {\rm M}_{\rm
p}^{\rm 5N - 4}} {\rm S}_2^{\rm 3N+1} \left ( \phi_1 \phi_2 \right )^{\rm N-1}
+
{\lambda_3 \over {\rm M}_{\rm p}^{\rm 8N-3}} \left ( \phi_1 \phi_2 \right
)^{\rm N-1} \left (
\phi_3 \phi_4 \right ) ^{3{\rm N}+1} + \cdots $$
$${\rm K} = {1 \over 2} \phi_{\rm i}^+ \phi_{\rm i} + {{\rm K_{34}^{S_2}} \over
{\rm M_p}} \left ( {\rm S}_2^+ \phi_3 \phi_4 + {\rm S}_2 \phi_3^+ \phi_4^+
\right ) + {{\rm K}_{12}^{12} \over {\rm M_p^2}} \left ( \phi_1^+ \phi_1
\phi_2^+ \phi_2 + \cdots \right )  + \cdots  \eqno(9)$$

\noindent where $\phi_{\rm i}$ is the set of all chiral superfields and ${\rm
M_p}$ is the Planck mass scale. For simplicity we take the gauge kinetic
function f = 1. The PQ symmetry is defined by the first term in W and the
second one in K being described by the charges
$${\rm R}_1 = {\rm R}_2 =  - 3 \hskip 1 truecm {\rm R}_3 = {\rm R}_4 =  1
\hskip 1 truecm {\rm R_S}_1 = 6 \hskip 1 truecm {\rm R_S}_2 = - 2 \ \ \ .
\eqno(10)$$

\noindent The lowest dimensional terms in the scalar potential breaking the PQ
symmetry have dimension 5N and are therefore sufficiently suppressed. Here the
solution ${\rm X}_1 + {\rm X}_2 = 0$ does not break PQ as in the previously
${\rm
E}_6 \times {\rm U}(1)_{\rm X}$ model. \par
Consider now the ${U(1)^2} \times U(1)_R$ axial symmetries of the model
defined in eq.(7) and (9). Denoting the corresponding parameters by
$\alpha$, $\beta$ and $\delta$, the fields transform as follows:

$$ \matrix{
\theta' & =\ &e^{-3i \delta \over 2} \theta \cr
{{\Phi'}_{1,2}}(\theta') & =\ &e^{i \alpha} \Phi_{1,2} (\theta) \cr
{{\Phi'}_{3,4}}(\theta') & =\ &e^{i \beta} \Phi_{3,4} (\theta) \cr
{S_{1}'}(\theta') &=\ &e^{-2i \alpha} S_{1} (\theta) \cr
{S_{2}'}(\theta') &=\ &e^{2i \beta} S_{2} (\theta).} \eqno(11)$$

All these symmetries are anomalous and the variation of the Lagrangian
gives
$$\eqalignno{
&\delta {\cal L}  \propto  \hskip 0.1cm [\alpha +{ \delta \over 2} +
3 (\beta +{\delta \over 2}) - 3{\delta \over 2} N] (F F^{*})_N +
[N (\beta + {\delta \over 2}) + l.e.] (F F^{*})_3 +
[N (X^2_1 + X^2_2) (\alpha + {\delta \over 2})  \cr
&+3 N (X^2_3 + X^2_4) (\beta +{\delta \over
2}) + X^2_5 (-2 \alpha + {\delta \over 2}) + X^2_6 (2 \beta +{\delta
\over 2})] (F F^{*})_{X} .&(12)} $$

  where $l.e.$ is the contribution of $(F F^{*})_3$ to the anomalies of
the axial symmetries coming from the low-energy sector. But the low-
energy color anomalous axial symmetries are forbidden due to the
Weinberg-Wilczek axion [3] which is experimentally excluded, so this
contribution vanishes. Then  to get a nonanomalous symmetry we must
separately put to zero the contribution to $\delta {\cal L}$ of the three
gauge groups $SU(N)\times SU(3)_c \times U(1)_X$ ,  but the only
solution of the three equations is the trivial one. So we have no
nonanomalous axial symmetry.

To check that at $\Lambda_{PQ}$ susy is dynamically broken we consider
the v.e.v. of the auxiliary component ${\rm F_S}_2$ given by
$${\rm F_S}_2 = - \left ( {\rm K}^{-1}  \right ) _{\rm S_2}^{\rm
S_2}{{\rm K}_{\rm S}}_2^{34} <\psi_3 \psi_4> .\eqno(13)$$

In the local supersymmetry case $K$ will be replaced by $G$, where [16]
$$ {\rm G} = {\rm K} + \log |W|^2 .\eqno(14) $$

\noindent  In this case the gravitino mass will have usual values ${\rm
m}_{3/2} \sim
{\Lambda_{\rm PQ}^3 \over {\rm M_p}^2}$. At the tree level in the supergravity
lagrangian soft scalar masses and couplings are generated through terms of
style
$${\rm R}_{\rm ijke} \left ( \psi^{\rm i} \psi^{\rm j} \right) \left (
\bar \psi^{\rm k} \bar \psi^{\rm l} \right ) .\eqno(15)$$

\noindent where ${\rm R}_{\rm ijkl} = \partial_{\rm i} \partial_{\rm j}
{\rm g}_{\rm kl} - {\rm g}^{\rm mn} {\rm g}_{\rm in,k} {\rm g}_{\rm mj,l}$ is
the curvature in the Kahler space and ${\rm g}_{\rm ij} = {\partial^2{\rm K}
\over \partial {\rm z}_{\rm i} \partial {\rm z}_{\rm j}^{*}}$. Actually
the soft masses are of order ${\Lambda_{\rm PQ}^3 \over {\rm M}_{\rm p}^2}$ so
larger values of $\Lambda_{\rm PQ}$ are preferred by this scenario. This is
allowed by the PQ breaking operator of dimension 5N in the scalar potential.
Gaugino masses will be generated at one-loop level due to the breaking of R
symmetry but will have rather small values. The interesting property of
this model is that it have no continuous nonanomalous $R$ symmetry and
nevertheless supersymmetry is broken. \par
To conclude, we can suppress the Planck scale effects and protect the PQ
symmetry in supersymmetric theories with an abelian ${\rm U}(1)_{\rm X}$
gauge group. The superfield content must be such that imposing the anomaly
cancellation and a single renormalisable term in the superpotential we can
have nontrivial abelian charges that forbid low dimensional operators breaking
the PQ symmetry. It is easy to construct models where no polynomial
nonrenormalisable superpotential can be written at all and Planck scale effects
vanish identically. \par
   After completion of this paper I became aware of the ref.[17] where
models with dynamical supersymmetry breaking and no $R$ symmetry are
constructed.
\vskip 0.5cm
\beginsection{\hbox {\bf Aknowledgments}}

I would like to thank Pierre Bin\'etruy and Carlos Savoy for useful
discussions and Jihad Mourad for a careful reading of the manuscript.
\vfill\eject
\centerline{\bf References} \bigskip \item{[1]} C. G. Callan Jr., R. F. Dashen
and D. J. Gross, Phys. Lett. {B63} (1976) 334; \item{} R. Jackiw and
C. Rebbi, Phys. Rev. Lett. {37} (1976) 172; \item{} G.'t Hooft, Phys.
Rev. {D14} (1976) 3432.  \item{[2]} R. D. Peccei and H. R. Quinn,
Phys. Rev. Lett. {38} (1977) 1440 ; Phys. Rev. {D16}
(1977) 1791.  \item{[3]} S. Weinberg, Phys. Rev. Lett. {40} (1978)
223; \item{} F. Wilczek, Phys. Rev. Lett. {40} (1978) 279.
\item{[4]} See e.g. M. S. Turner, Phys. Rep. {197} (1991) 67.
\item{[5]} J. E. Kim, Phys. Rev. Lett. 43 (1979) 103; \item{} M.
A. Shifman, A. I. Vainshtein and V. I. Zakharov, Nucl. Phys. {B166}
(1980) 493; \item{} M. Dine, W. Fischler and M. Srednicki, Phys. Lett.
{B104} (1981) 199.   \item{[6]} J. E. Kim, Phys. Rev.
{D31} (1985) 1733; \item{} D. B. Kaplan, Nucl. Phys.
{B260} (1985) 215.  \item{[7]} A. Nelson, Phys. Lett.
{B136} (1984) 387; \item{} S. M. Barr, Phys. Rev. Lett.
{53} (1984) 329. \item{[8]} L. Bento, G. C. Branco and P. A. Parada,
Phys. Lett. {B267} (1991) 95. \item{[9]} R. Holman, S. Hsu, T.
Kephart, E. Kolb, R. Watkins and L. Widrow, Phys. Lett. {B282}
(1992) 132. \item{[10]} M. Kamionkowski and J. March-Russell, Phys. Lett.
{B282} (1992) 137; \item{} S. Barr and D. Seckel, Bartol preprint.
\item{[11]} L. Randall, Phys. Lett. {B284} (1992) 77.
\item{[12]} H. Georgi, L. J. Hall and M. B. Wise, Nucl. Phys.
{B192} (1981) 409.
\item{[13]} See e.g. H. P. Nilles, Phys. Rep. {110} (1984) 1.
\item{[14]} G. Veneziano and S. Yankielowicz, Phys. Lett. B113 (1982) 231;
T. R. Taylor, G. Veneziano and S. Yankielowicz, Nucl. Phys. B218(1983) 493;
I. Affleck, M. Dine and N. Seiberg , Nucl. Phys. B241 (1984) 493 , Phys.
Lett. B140 (1984) 59;
Y. Meurice and G. Veneziano , Phys. Lett. B141 (1984) 69.
\item{[15]} I. Affleck, M. Dine and N. Seiberg, Nucl. Phys. {B256}
(1985) 557.
\item{[16]} E. Cremmer, S. Ferrara, L. Girardello and A. Van Proeyen ,
Nucl. Phys. {B212} (1983) 413; \item{} P. Bin\'etruy, G. Girardi and
R. Grimm, preprint LAPP-TH-275/90.
\item{[17]} A. E. Nelson and N. Seiberg, UCSD/PTH 93-27, hep-ph@xxx/9309299.
\bye